\documentstyle[sprocl,epsf]{article}

\def\ra{\rightarrow}

\def\be{\begin{equation}}
\def\ee{\end{equation}}
\def\bea{\begin{eqnarray}}
\def\eea{\end{eqnarray}}

\def\la{\langle}
\def\ra{\rangle}

\def\beq{\begin{eqnarray}}
\def\eeq{\end{eqnarray}}

\def\btem{\bibitem}

\begin{document}

\title{Precritical Chiral Fluctuations in Nuclei\footnote{Talk
 given at the 1998 YITP-workshop on QCD and Hadron Physics
(Oct. 14-18, 1998, Kyoto, Japan).  Extended 
 version of this talk is given  in nucl/th-9810022 
 (T. Hatsuda, T. Kunihiro  and H. Shimizu)}}

\author{T. HATSUDA}

\address{
Physics Department, Kyoto University, \\ Kyoto 606-8502, Japan \\
E-mail: hatsuda@ruby.scphys.kyoto-u.ac.jp} 

\author{T. KUNIHIRO}

\address{
Faculty of Science and Technology, Ryukoku
 University,\\ Seta, Otsu-city, 520-2194, Japan\\
E-mail: kuni@math.ryukoku.ac.jp}


\maketitle\abstracts{ 
 Spectral enhancement  near 2$m_{\pi}$ threshold 
 in the $I$=$J$=0 channel in nuclei
 is shown to be a distinct  signal
 of the partial restoration of chiral symmetry.
 The relevance of this phenomenon 
 with possible detection of 
 $2\pi^{0}$ and $2\gamma$ 
 in hadron-nucleus and photo-nucleus reactions
 is discussed.}

\section{Introduction}
 The low density theorem in QCD and  model calculations 
 suggest that partial restoration of chiral symmetry takes place
 in nuclear medium.\cite{HK94}
  If the quark condensate
 decreases substantially
 in nuclear matter, 
  fluctuation of the condensate becomes 
  large in accordance
 with the general wisdom of statistical physics.\cite{GK}
 This implies that there is a softening of a collective excitation 
 in  the scalar-isoscalar channel with the following
  physical consequences \cite{HK85}:
 (i)  partial degeneracy of the scalar-isoscalar particle
 (traditionally called the $\sigma$-meson) with the pion, and
 (ii) decrease of the decay width of $\sigma$
  due to  the phase space suppression caused by (i) in the reaction
 $\sigma \rightarrow 2 \pi$.
  Although $\sigma$ appears only as a broad resonance in the vacuum
 and is hard to be distinguished from the 
 background \cite{pipi},
 it was suggested that it may 
 appear as a sharp resonance at finite temperature ($T)$ because
 of (i) and (ii) discussed above. \cite{HK85,later,CH98}

  In this talk, we demonstrate, using a toy model,
  that the spectral enhancement
  associated with the partial chiral restoration  
  takes place also at finite baryon density close to 
  $\rho_0 = 0.17 {\rm fm}^{-3}$.
   As possible experiments
 to detect this softening,
 we will discuss the production of the neutral-dipion (2$\pi^0$)
 and diphoton (2$\gamma$)  in reactions with 
   heavy nuclear targets.
  We also mention the relevance of the softening and
  the recent  measurement of the near-threshold
 $\pi^{+}\pi^{-}$ production in $\pi^{+}$-nucleus reactions.
 \cite{ppp}

\section{Basic idea}

 Let us first
 describe the general
 aspects of the spectral enhancement near the two-pion threshold.
 Consider the propagator 
 of the $\sigma$-meson at rest in the medium :
$ D^{-1}_{\sigma} (\omega)= \omega^2 - m_{\sigma}^2$
$ - \Sigma_{\sigma}(\omega;\rho)$,
where $m_{\sigma}$ is the mass of $\sigma$ in the tree-level, and
$\Sigma_{\sigma}(\omega;\rho)$ is 
the loop corrections
 in the vacuum as well as in the medium.
 The corresponding spectral function reads
\beq
\label{spect1}
\rho_{\sigma}(\omega) 
 = - {1 \over \pi} 
{ {\rm Im} \Sigma_{\sigma} \over (\omega^2 - m_{\sigma}^2 -
 {\rm Re} \Sigma_{\sigma} )^2
  + ({\rm Im} \Sigma_{\sigma} )^2 }.
\eeq
Near the two-pion threshold, the phase space factor gives
${\rm Im} \Sigma_{\sigma} \propto \theta(\omega - 2 m_{\pi}) \ 
	 (1 - 4m_{\pi}^2 / \omega^2)^{1/2} $ in 
the one-loop order.
 On the other hand, partial restoration of chiral
 symmetry indicates that $m_{\sigma}^*$ (``effective mass'' of $\sigma$ 
 defined by 
 ${\rm Re}D_{\sigma}^{-1}(\omega = m_{\sigma}^*)=0$)
  approaches to $ m_{\pi}$.  Therefore,
 there exists a density $\rho_c$ at which 
 ${\rm Re} D_{\sigma}^{-1}(\omega = 2m_{\pi})$
 vanishes even before the complete $\sigma$-$\pi$
 degeneracy takes place.
At $\rho = \rho_c$, the spectral function is solely dictated by the
 imaginary part of the self-energy;
\beq
\rho_{\sigma} (\omega \simeq 2 m_{\pi}) 
 =  - [\pi \ {\rm Im}\Sigma_{\sigma} ]^{-1}
 \propto \theta(\omega - 2 m_{\pi}) \ 
 (1-{4m_{\pi}^2 \over \omega^2 })^{-1/2}.
\eeq
This implies that, even if there is no sharp resonance
 in the scalar channel in the vacuum, there arises
  a mild (integrable) singularity just above the threshold in the medium.
 This is a general phenomenon associated with the 
 partial restoration of chiral symmetry.

\section{Linear sigma model}

 Let us now evaluate $\rho_{\sigma}(\omega)$ in a {\em toy} model, namely
  the SU(2) linear $\sigma$-model:
\beq
\label{model-l}
{\cal L} & = &  {1 \over 4} {\rm tr} [\partial M \partial M^{\dagger}
 - \mu^2 M M^{\dagger} 
  - {2 \lambda \over 4! } (M M^{\dagger})^2   - h (M+M^{\dagger}) ],
\eeq
where tr is for the flavor index and  
 $M = \sigma + i \vec{\tau}\cdot \vec{\pi}$.
 Although the model 
 is not a precise low energy representation of QCD, \cite{GL} it
 is known to describe the
 pion dynamics qualitatively well up to 1GeV. \cite{BW} 

 The coupling constants $\mu^2, \lambda$ and $h$, which
 are determined in the
 vacuum to reproduce $f_{\pi}=93$ MeV, $m_{\pi}=140$ MeV as well as
 the s-wave $\pi$-$\pi$ scattering phase shift in the one-loop order,
 are recapitulated for two 
 characteristic cases in Table 1 in which
  $m_{\sigma}^{peak}$ is defined as 
  a peak position of $\rho_{\sigma}(\omega)$.

\begin{table}[t]
\caption{Parameters for $m_{\sigma}^{peak}=$ 550, 750 MeV
taken from ref.6. \label{tab1}} 
\vspace{0.2cm}
\begin{center}
\footnotesize
\begin{tabular}{|c|c|ccc|} 
\hline
                            &
$m_{\sigma}^{peak}$ (MeV) \ \ &
$\sqrt{-\mu^2}$ (MeV)         &
$\lambda/4 \pi$              & 
$h^{1/3}$  (MeV)                  \ \\ \hline
(I)  &  550 & 284 & 5.81  & 123   \\
(II) &  750 & 375 & 9.71  & 124   \\
\hline
\end{tabular}
\end{center}
\end{table}
%

Let us now consider the $\sigma$-meson embedded in nuclear matter.
 The interaction of $M$ with the nucleon  $N$
 in SU(2) chiral symmetry  is modeled as
\beq
\label{int-nm}
{\cal L}_{I}(N, M) =
  - g \chi \bar{N} U_5 N  - m_0  \bar{N} U_5 N ,
\eeq
where we have used a polar representation 
 $ \sigma + i  \vec{\tau}\cdot \vec{\pi} \gamma_5 \equiv \chi U_5 $ 
 for convenience. 
 The first term in (\ref{int-nm}) with  $g$
 is a standard chiral invariant coupling in the linear $\sigma$
 model.  The second term with a new parameter $m_0$ 
 is also chiral invariant and non-singular, but is 
  not usually taken into account in the literatures.

 After the dynamical breaking of chiral symmetry
 in the vacuum ($\langle \sigma \rangle_{\rm vac} \equiv \sigma_0 \neq 0$), 
 the coupling of  $\tilde{\sigma} = \sigma - \sigma_0$ and  $\pi $
 with $N$ is dictated by 
 $g_{\rm s} \equiv g$ and $g_{\rm p} \equiv g_{\rm s} + m_0 /  \sigma_0)$,
 respectively.
   Because of $m_0$,
 the standard constraint $g_s = g_p$ can be relaxed
 without conflicting with chiral symmetry.
  $g_{\rm p}$ is constrained as 
 $g_{\rm p} = m_N /  \sigma_0 \simeq m_N/f_{\pi}
 = g_{\pi N} = 13.5$, while
 $g_{\rm s}$ is independent from $g_{\rm p}$
 and can be treated as a free parameter.
 This freedom partially avoid the well-known
 problem  that 
  $g_s=g_p$ combined with eq.(\ref{model-l})
 does not reproduce the known nuclear
 matter properties in the mean-field level. \cite{sw}

\section{One-loop estimate}

Let us parametrize the chiral condensate
 in the medium as 
\beq
\label{cond}
\langle \sigma \rangle \equiv  \sigma_0 \ \Phi(\rho).
\eeq
In the linear density approximation,
 $\Phi(\rho) = 1 - C \rho / \rho_0 $ with
 $C = (g_{\rm s} /\sigma_0 m_{\sigma}^2) \rho_0$.
  The plausible value of $\Phi(\rho = \rho_0)$ is
 0.7 $\sim$ 0.9. \cite{HK94}

The one-loop corrections to the self-energy for $\sigma$ can be read off
 from the  diagrams in Fig.1:
$\Sigma_{\sigma}(\omega;\rho) = \Sigma_{\rm vac}^{A} $
$ + \Sigma_{\rm vac}^B + \Sigma_{MF}(\rho) + \Sigma_{ph}(\rho)$.
 $\Sigma_{\rm vac}^A$ ($\Sigma_{\rm vac}^B$) corresponds to
 Fig.1A and Fig.1B , respectively. 
  The explicit form of 
  $\Sigma_{\rm vac}^{A+B}$  
 is given in Appendix A of ref.6.
  $\Sigma_{MF}(\rho)$ corresponds to the mean field correction
 in the nuclear matter (Fig.1C).
 $\Sigma_{ph}(\rho)$ is a correction from the nuclear particle-hole 
 excitation. We take only the density dependent part
 in these diagrams and
 neglect the problematic vacuum-loops of the nucleon. 
 \cite{sw}

\begin{figure}[t]
\centerline{
    \epsfxsize=8cm
    \epsfbox{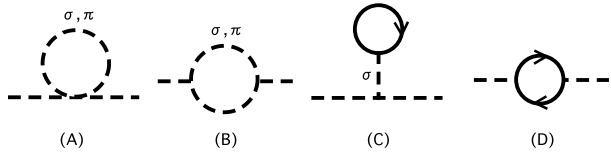}
}

\caption{One-loop self energy.
 The dashed (solid) line denotes $\sigma$ or $\pi$ (N).}
 \label{fig1}
\end{figure}

The leading term in the mean-field part is easily estimated as
\beq
\Sigma_{MF}(\rho) =  \lambda \sigma_0 \ (\la \sigma \ra - \sigma_0)
 = - \lambda \sigma_0^2 \ (1-\Phi(\rho)) ,
\eeq
 Leading term  in the particle-hole part (Fig.1D)
  starts from $O(\rho^{5/3})$
  and is not more than a few \% of $\Sigma_{MF}(\rho)$
  at  $\rho = \rho_0$.
 This is in contrast to the case of the pion, where
 both $\Sigma_{MF}(\rho) $ and $\Sigma_{ph}(\rho) $ are
 proportional to $\rho$ and cancel with each other due to chiral symmetry.  
 Because of this cancellation, we can neglect the 
 two-loop contribution related to the medium modification
 of  the low-momentum pion  near
 the two-pion threshold.

 Up to the order we are considering, 
 ${\rm Im} \Sigma_{\sigma}$  solely comes from 
${\rm Im} \Sigma_{\rm vac}^B$, since 
 there is no Landau damping and scalar-vector
 mixing for the $\sigma$-meson at rest in nuclear matter. 
 For $2 m_{\pi} \le \omega \le 2m_{\sigma}$,
\beq
\label{vac4}
{\rm Im} \Sigma_{\sigma}(\omega;\rho) =
{\rm Im} \Sigma_{\rm vac}^{B}
 = - {\lambda^2 \over 96 \pi}  \sigma_0^2
 \sqrt{1 - {4m_{0\pi}^2 \over \omega^2}}  .
\eeq

As we have already discussed,
the threshold peak is expected to be prominent when
  ${\rm Re} D_{\sigma}^{-1} = \omega^2 - m_{ \sigma}^2 -
 {\rm Re} \Sigma_{\sigma}  =0 $. 
 In terms of $\Phi$,
 this condition is  rewritten as
 $\Phi(\rho_c) = 0.74$ \ \ \ (for case  (I) in table 1), and 
    \  $0.76$  \ \ \ (for case  (II) in Table 1).
 The numbers in the right hand side are rather insensitive to the
 parameters  in Table 1
 as far as the physical quantities in the vacuum 
 are fixed.
 In the linear density formula $\Phi(\rho) \simeq 1- 0.2 \rho/\rho_0$, 
 we obtain $\rho_c \simeq 1.25 \rho_0$.
 The spectral functions 
 together with ${\rm Re} D_{\sigma}^{-1}(\omega)$  
 for two cases (I) and (II) 
 are shown in Fig.2. 
   The characteristic enhancements 
 just above the 2$m_{\pi}$  threshold are seen for
  $\rho \simeq \rho_c$.


\begin{figure}[t]
\centerline{
    \epsfysize=5.7cm
    \epsfxsize=5.7cm
    \epsfbox{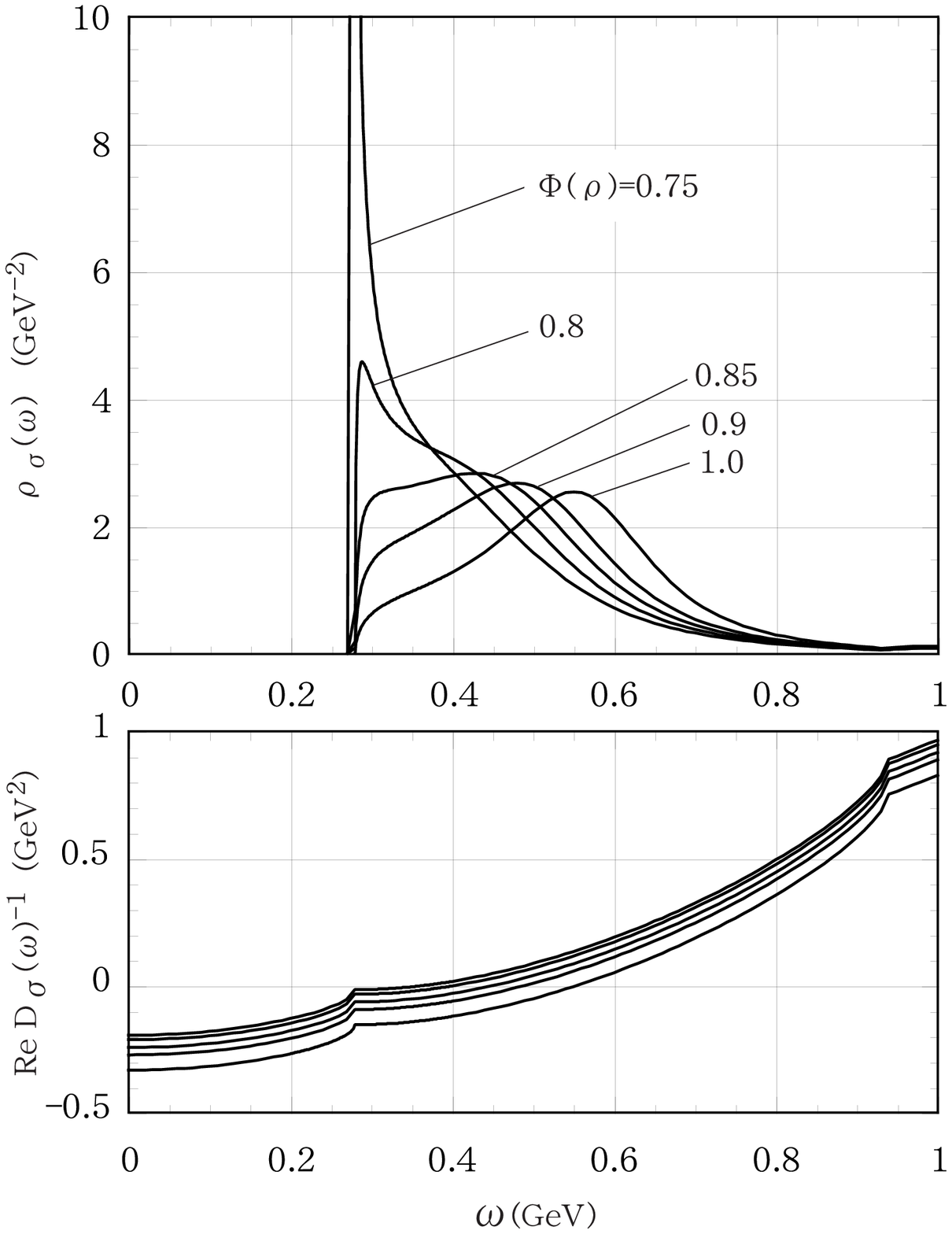}\hspace{0.2cm}
    \epsfysize=5.7cm
    \epsfxsize=5.7cm
    \epsfbox{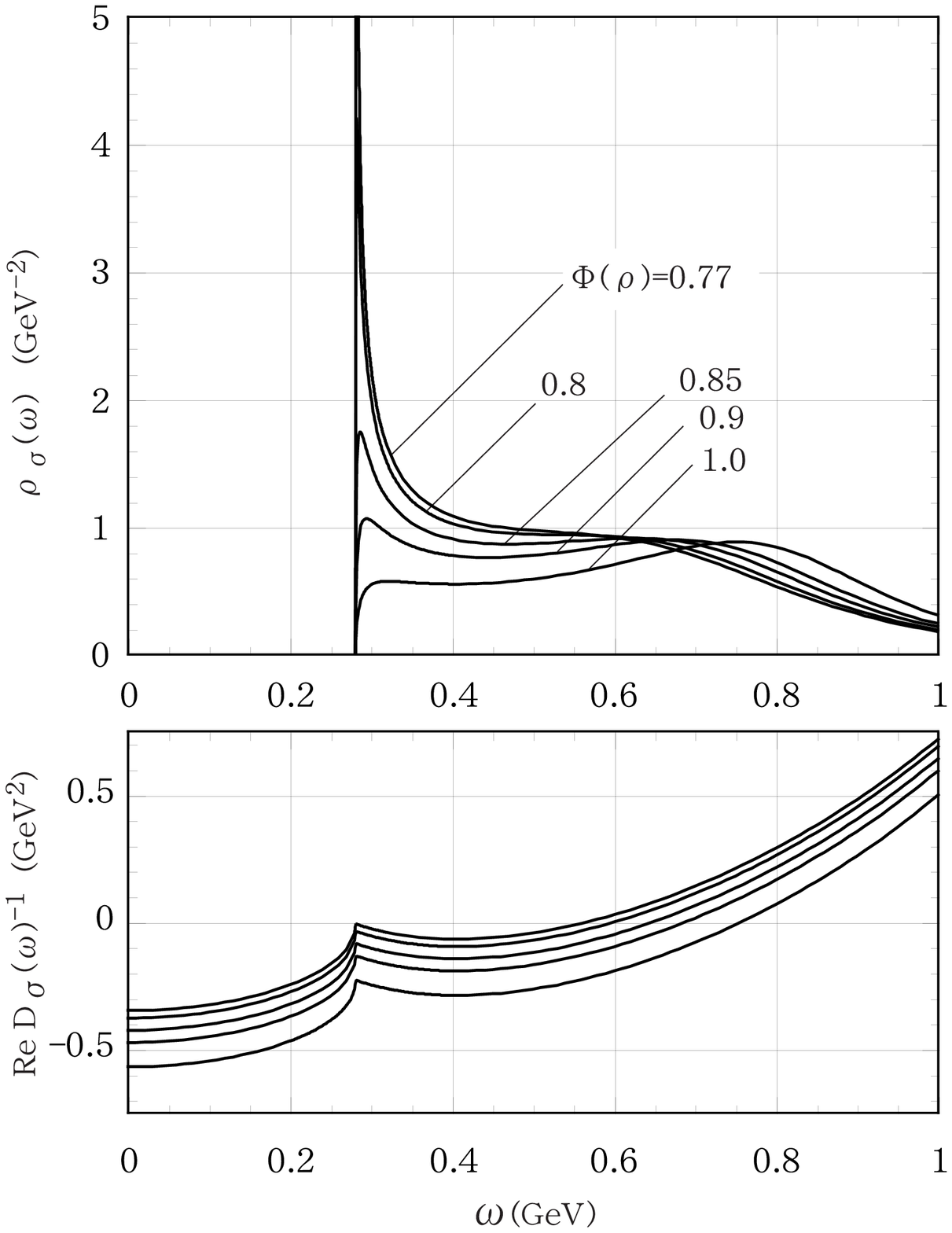}
}

\caption{Left: Spectral function for $\sigma$ and  the 
 real part of $D_{\sigma}^{-1}$  for several values of
 $\Phi = \la \sigma \ra / \sigma_0$ with $m_{\sigma}^{peak}
 = 550$ MeV (case (I) in Table 1). In the lower panel,
 $\Phi$ decreases from bottom to top. Right: Same 
for $m_{\sigma}^{peak}= 750$ MeV (case (II) in Table 1.)  }
\label{fig2}
\end{figure}


  We emphasize here that 
 there are two reason behind the enhancement: partial restoration
 of chiral symmetry ($m_{\sigma}^* \rightarrow 
  m_{\pi}$),  
 and the cusp structure of ${\rm Re}D^{-1}(\omega = 2 m_{\pi}$.
 Both features can be seen in the lower panels of Fig.2:
 Although the cusp is not prominent at zero density,
 it eventually hits the real axis at $\rho = \rho_c$
 because ${\rm Re}D^{-1}(\omega )$ increases associated
 with $m_{\sigma}^* \rightarrow 2 m_{\pi}$.
  This is a general phenomena for systems where the 
 internal symmetry is partially restored in the medium.
   Another important observation is that, 
  even at  densities 
 well below the point where $m_{\sigma}^*$ and $m_{\pi}$ are
 degenerate,
 one can expect a large enhancement of 
 $\rho_{\sigma}(\omega \simeq 2m_{\pi})$.

\section{Experiments}

To confirm the threshold enhancement associated with the
 partial chiral restoration, measuring 2$\pi^0$ and 
$2\gamma$ with hadron/photon beams off
 the  heavy nuclear targets should be most appropriate. 
 By measuring $\sigma \rightarrow 2 \pi^0 \rightarrow
  4\gamma$, \cite{4gamma}
 one can avoid the possible $I=J=1$ background from the
 $\rho$ meson inherent in the $\pi^+\pi^-$ measurement.
 Measuring the electromagnetic decay $\sigma \rightarrow 2 \gamma$
 is also important because of the small final state
 interactions. \cite{HK85} 
 There is also a possibility that one can detect dilepton 
 through the scalar-vector mixing in matter: $\sigma \to \gamma^* \to
 e^+ e^-$. \cite{lepton}

 To enhance the production cross section of the 
 critical fluctuation in the $\sigma$-channel,
   (d, $^3$He)  reactions may be useful as the
 case of $\eta$ and $\omega$ mesic nuclei.\cite{HHG}
  To cover the spectral function 
 in the range  $2m_{\pi} < \omega < 750$ MeV,
 the incident kinetic energies of $d$
  in the laboratory system is estimated as
  $1.1 {\rm GeV} < E < 10$ GeV,

Recently  CHAOS collaboration  \cite{ppp} measured the 
$\pi^{+}\pi^{\pm}$
invariant mass distribution $M^A_{\pi^{+}\pi^{\pm}}$ in the
 reaction $A(\pi^+, \pi^{+}\pi^{\pm})X$ with the 
 mass number $A$ ranging
 from 2 to 208: They observed that
the   yield for  $M^A_{\pi^{+}\pi^{-}}$ 
 near the 2$m_{\pi}$ threshold is close to zero 
for $A=2$, but increases dramatically with increasing $A$. They
identified that the $\pi^{+}\pi^{-}$ pairs in this range of
 $M^A_{\pi^{+}\pi^{-}}$ is in the $I=J=0$ state.
  Attempts so far in hadronic models without considering the
 partial chiral restoration 
 failed to reproduce this enhancement. \cite{wambach,oset}
 On the other hand, $A$ dependence of the 
 the invariant mass distribution of the CHAOS data
 near 2$m_{\pi}$ threshold has close
 resemblance to our model calculation in Fig.2, which suggests
 that this experiment may already provide
  a hint about how  the (partial) restoration of chiral symmetry
 manifests  itself  at finite density.

\section{Summary}

In summary,  we have shown that the spectral function in the
 $I=J=0$ channel has 
 a large enhancement near the $2m_{\pi}$ threshold even at
  nuclear matter density due to the partial chiral restoration.
  Detection of the 
 dipion and diphoton spectral distribution
 in the reactions of hadron/photon with heavy nucleus is suitable
  to confirm the idea of partial chiral restoration in nuclei.

\end{document}